# Towards Mixed Reality as the Everyday Computing Paradigm: Challenges & Design Recommendations

Amir Reza Asadi & Reza Hemadi

Humind Lab, Iran[1]

*Abstract*— This research presents a proof-of-concept prototype of an all-in-one mixed reality application platform, developed to investigate the needs and expectations of users from mixed reality systems. The study involved an extensive user study with 1,052 participants, including the collection of diaries from 6 users and conducting interviews with 15 participants to gain deeper insights into their experiences. The findings from the interviews revealed that directly porting current user flows into 3D environments was not well-received by the target users. Instead, users expressed a clear preference for alternative 3D interactions along with the continued use of 2D interfaces. This study provides insights for understanding user preferences and interactions in mixed reality systems, and design recommendations to facilitate the mass adoption of MR systems.

*Keywords—Mixed Reality, AR, VR, 3D Interfaces*

## I. Introduction

The dawn of Mixed Reality (MR) beckons us towards a new era of human-computer interaction, where the boundaries between the physical and digital realms begin to blur. With the increasing ubiquity of MR devices and applications, it becomes imperative to explore the potential impact of this technology on our everyday lives and its transformative role in shaping our computing experiences.

Mixed reality represents a spectrum of immersive environments, spanning from the real world to fully virtual domains [Milgram]. Augmented reality and augmented virtuality applications find their place along this spectrum, offering varying levels of immersion and interaction.

For the industry to advance towards widespread adoption of these systems, it becomes imperative to address the challenges and pain points associated with their everyday use. One such concern is the potential cognitive load and distraction caused by the presentation of virtual information, necessitating the design of interactions that remain unobtrusive and contextually relevant [Bowman][Grubert]. Moreover, transitioning between different points on this immersive spectrum may further exacerbate these issues, warranting careful consideration in design and implementation.

Mixed reality (MR) systems possess the remarkable capability of extending the boundaries of interaction beyond conventional screens, effectively leveraging the spatiality of users' environments Consequently, the design of MR systems necessitates a thorough exploration of 3D interfaces to harness their full potential Traditionally, 3D interfaces were predominantly associated with video games, while interaction designers tended to steer clear of integrating them into mass target use cases due to concerns surrounding the additional cognitive load imposed on users [LaViola Jr][Kyritsis].

In this research endeavor, we delved into the usability aspects of Mixed Reality (MR) environments, encompassing both 3D and 2D interfaces, along with natural language processing, in the context of everyday interactions. The study focused on observing users' interactions across various scenarios, including social interaction, productivity, collaboration, and playfulness. Notably, the experiment uniquely leveraged users' interactions within their natural settings, departing from the confines of the lab setting. This approach allowed us to uncover valuable insights into user expectations and preferences, culminating in a set of design implications tailored for crafting mixed reality interactions.

## II. Research Methodlogy

The study started by envisioning a mixed reality application specifically tailored for everyday interactions with information across various use cases. To achieve this objective, a cohort of 6 participants (comprising 3 males and 3 females) was carefully selected to explore and understand their daily information interactions. Based on the insights garnered from this investigation, a functional prototype of a social platform for mixed reality information interaction was meticulously crafted.

The application development targeted the iOS platform, as it offered a larger user base for testing purposes, and Apple's SDKs facilitated a more expedited development process.

 The aforementioned participants were also enlisted to partake in a diary study, allowing us to gain deeper insights into their usage patterns and experiences with the app in their everyday lives. Furthermore, we sought to expand the user base by sharing the app in online forums while respectfully requesting users' consent to collect data on their interactions with the system. This endeavor resulted in the collection of data from an impressive cohort of 1,052 users.

Concluding the study, 15 users were interviewed to elicit their expectations, thereby enriching our understanding of user preferences. These approaches collectively contributed to the comprehensive exploration of mixed reality interactions with information, laying the groundwork for potential enhancements and refinements in the realm of everyday computing.

Acknowledging the constraints posed by handheld devices, which restricted hands-free interactions in our mixed reality system, we proactively sought to enhance user experience. To address this limitation, we distributed 15 tripods to interview participants, empowering them with a hands-free setup and enabling easier interactions with the mixed reality system.

---

[1] The research was conducted in 2020-2022 in Humind Lab a private company.

### III. MULTIPURPOSE MIXED REALITY SYSTEM

In our endeavor to simulate the interactions of a Mixed Reality Operating System, we have developed an expansive iPhone application. As illustrated in Figure 1-2, this application grants users access to a diverse array of virtual functionalities, including a virtual items shop, two interactive games, spatial content authoring, socializing, messaging capabilities, a personalized virtual home portal, and interactive 3D Cork Boards that function as collaboration and office applications.

Within the application, users have the freedom to choose their preferred mode of interaction. They can navigate through the main application features using a flat screen menu or opt for a more immersive experience by engaging with a 3D virtual assistant. Leveraging the gyroscope of their device, users can also explore the exciting realm of 3D interaction, further enhancing their engagement with the simulated Mixed Reality Operating System.

Additionally, users can discover other platform users in their physical environment through 3D augmented profile banners. Furthermore, users have the unique capability to invite others from far distances to join them in their personalized virtual home, bridging the gap between physical and virtual realms through augmented virtuality. This dynamic feature elevates the social aspect of the application, fostering a sense of virtual presence within their surroundings, and forging connections beyond geographical constraints.

### IV. FINDINGS

In this section we provide findings about usage of this system.

#### A. Usage Data

An impressive insight from the testing data revealed the distinct levels of interest in different application features. A vast majority of testers(87%) demonstrated curiosity in the "add content" button, signaling a strong desire for content creation and contribution within the mixed reality environment.

Moreover, more than half of users participated in the available interactive games, whereas a little over a third of users engaged with the virtual shopping experience, indicating varying levels of engagement with the entertainment and e-commerce functionalities within the application.

Within the user population, a notable portion interacted with the personalized virtual home portal for customization purposes, while a smaller fraction made use of the social features. Additionally, a significant number of users utilized the interactive 3D Cork Boards as practical tools for collaboration and office-related tasks.

.

#### B. Frustrations

*1) User Interface Issues*

During our investigation, users encountered several significant challenges while interacting with the application, primarily centered around the stability of displaying virtual elements within their physical environment. This issue encompassed conflicts arising from object occlusion, wherein virtual objects interacted with real-world elements, disrupting the seamless blending of digital and physical realities. Additionally, integrating UI elements into the real-world environment further compounded the complexity of the user interface as testers grappled with the visual coherence between virtual and physical elements. While the Semi-transparent UI elements were used to increase the immersion, the dynamic nature of real world environments has decreased the visual contrast in some cases.

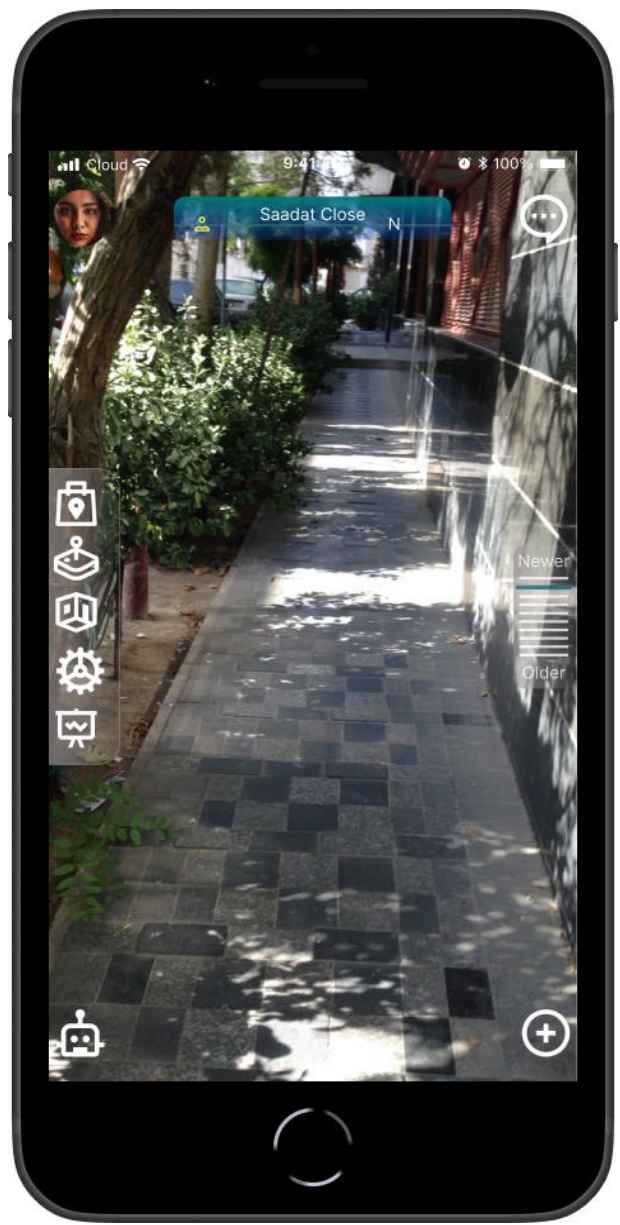

Fig 1. Mainpage UI

Remarkably, an intriguing pattern emerged during the testing phase. Participants who were not part of the diary studies demonstrated a lower propensity to explore various application features independently. Instead, they heavily relied on strong and directive visual cues presented within the app environment to guide their interactions.

*2) Lack of Established Mental Models*

for effectively utilizing mixed reality in their everyday lives. Users expressed uncertainty and hesitancy in integrating the application into their daily routines, indicating a need for clearer guidance on how to leverage mixed reality technology to its fullest potential.

The issue of physical movement emerged as a significant concern among users. While they acknowledged the benefits of a 3D environment for navigating information, the practical challenges of handheld augmented reality, which required physical movement to accomplish tasks, presented an obstacle. Users found that some traditional methods for completing tasks were more effective, leading to a disconnect between the promise of 3D immersion and the real-world constraints of physical movement.

*3) Integration Challenges with non AR/VR Workflows*

Moreover, the diary study participants, who exclusively focused on the application during the testing period, mainly engaged in writing spatial memos and accessing spatially generated content. However, they expressed uncertainty about how to make the application an integral part of their daily lives beyond the testing phase.

P2: *"The problem is that none of my daily workflow applications are compatible with it."*

*4) Need for More Immersive Content*

Participants who utilized the virtual home feature showed interest in customizing their virtual living spaces and sharing them with friends, indicating the potential appeal of social mixed reality and virtual environments. However, no user used this feature more than 3 times, and the lack of diverse 3D content was a recognized reason behind this behavior in interviews.

P9: *"There was no more virtual goods to decorate my room".*

*5) Need for More Content Generation Tools*

A notable theme that emerged from the interviews was the expressed need for greater customization options and personalization features.

P12: "I wish I could bring my statutes to my portal".

Users sought more control over the content generation process, reflecting their desire to tailor the mixed reality experience according to their individual preferences and tastes.

*C. Expectations from a System for Everyday Usage*

We posed the following questions to the interviewees, and after analyzing their responses, we coded their expectations as depicted in Figure 3.

- How would you prefer to launch the Portal?
- How would you prefer to start the games?
- How would you like to control the game?
- How would you change the process of running a feature?
- How would change the process of editing 3D Cork Board?
- How would you prefer to navigate the menus?

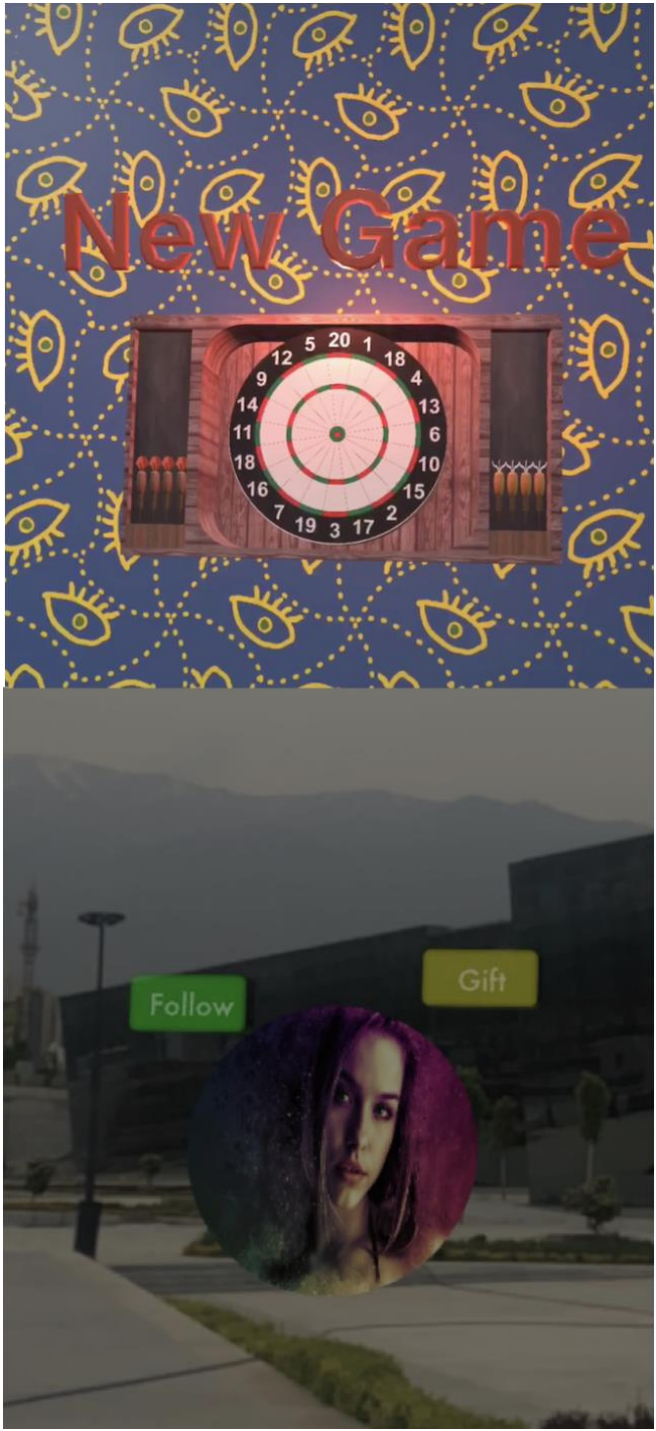

Fig 2

Top Photo: The image showcases the immersive 3D interface of a game within a user's virtual home. To initiate the game, the user gazes directly towards the New Game Button. Additionally, users have the option to personalize the wallpaper of their portal which may lead to a reduction in contrast between visual elements.

Bottom Photo: This image portrays a virtual representation of another user seamlessly integrated into the main user's environment. Users can opt to follow the virtual representation or send virtual gifts within the mixed reality setting.

Following the testing phase, we conducted comprehensive in-depth interviews to gain deeper insights into users' perspectives and expectations. Our findings prominently highlighted the lack of established mental models among users

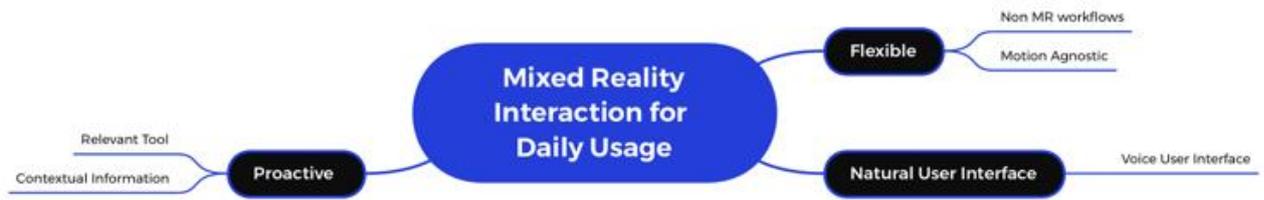

- Please describe your ideal way of engaging with virtual content and objects.
- How would you prefer to navigate the immersive environment?
- What is missing in this system for using it every day?

The analysis of these answers is depicted in Figure 3, as described in the section below.

*1) Proactive User Flows*

Participants expressed an expectation that the system would adapt its behavior according to their contextual situations. They anticipated that the system would dynamically change its interface based on the users' status, such as whether they are walking or standing, to deliver a more personalized and contextually relevant experience. For instance, they envisioned the system showing relevant tools and contextual information that align with their current activity or location.

Participant 3 shared, "The system can display walking status and locational information while I'm moving, and when I'm in my bed, it can show the game icons." This statement highlights the desire for the system to provide real-time updates on walking status and location while in motion, while also transitioning to show entertainment-related icons when the user is stationary in their bed.

By adapting to the users' specific contexts, the system aims to enhance user engagement and efficiency by presenting them with relevant information and tools based on their current activities and surroundings.

*2) Natural User Interface*

Another expectation expressed by participants was the use of a voice user interface. Participants 5, 6, and 13 mentioned that they preferred describing items while interacting with the 3D cork board, believing it would be easier for everyday use. Using voice commands allowed them to effortlessly manipulate the board's content, making it a more convenient and natural interaction method.
Additionally, participants 12 and 15 showed interest in using voice commands for calling virtual objects in games instead of selecting items from menus. They believed that this approach would offer a more user-friendly experience, making the gaming interactions smoother

*3) Flexibility*

A significant portion of the interviewees who used the 3D Corkboard for document authoring expressed a preference for using 2D user interfaces (UI) over 3D interactions for this specific task. They also conveyed an interest in using the 3D Mixed Reality Environment primarily for viewing documents. Furthermore, some testers mentioned that they faced difficulties using 3D interactions while napping and expressed the need for a motion-agnostic alternative for those instances when they were using the apps in a reclined position. This observation was also evident in the user screenshots. Additionally, we noticed that 35% of users installed the app while they were resting on a bed or sofa. This finding, combined with the observed lack of usability at the beginning of their experience, seemed to have a significant impact on their overall perception of mixed reality.

Fig 3. Expectations from MR systems for Daily Usage Mind map

In general, users expressed a strong expectation for flexibility in achieving tasks through multiple methods while using the system in their everyday lives.

By providing users with a choice between 2D and 3D interactions and addressing the challenges related to specific situations, the system can better cater to users' preferences and practical needs for daily use. The focus on adaptability and user-centered design can enhance the overall user experience, making the mixed reality system more versatile and user-friendly.

## V. DESIGN RECOMMENDATIONS

In this section, we have developed design recommendations that will guide our future work and implementation.

*A. Mixed Reality Action Store , Instead of Application Store*

To address users' needs and develop proactive user flows, a practical approach involves focusing on developing actions instead of traditional applications. By enabling users to perform specific actions in various situations, such as seamlessly shifting their focus to entertaining activities or accomplishing professional tasks, we can create a more dynamic and engaging user experience. We are actively working on a conceptual framework to expand this mindset and incorporate it into our design approach. This framework aims to provide software designers with the tools and mindset to tailor their solutions and meet users' expectations for a proactive user experience. By embracing this approach, we strive to create a more intuitive and seamless interaction model, enhancing users' overall satisfaction and engagement in mixed reality environments.

Embracing this action-centric approach can also facilitate a context-aware paradigm by creating a grounding for mapping user intentions with relevant actions to their context. This contextual alignment has the potential to enhance the overall user experience, making interactions more intuitive and relevant in mixed reality environments.

*B. 2D to 3D Transitions*

In the preceding section, we gained insights into the user experience, indicating that an interface allowing users to seamlessly utilize both 2D and 3D elements simultaneously could be ideal. To facilitate smooth transitions between these modes and accommodate users' preferences, we focused on developing user flows and interactions that support both approaches concurrently. As a result, we devised tri-fold 3D documents, which enhance the readability of documents within the 3D environment. Simultaneously, users can effortlessly switch to a 2D single-page view, ensuring a versatile and adaptable reading experience.

*C. Context Aware Dynamic User Interface*

Just as responsive and adaptive web design revolutionized the usability of websites across various screen sizes, the widespread adoption of mixed reality in daily life demands a similarly versatile approach. A context-aware dynamic user interface (UI) becomes paramount in creating a seamless and user-friendly mixed reality experience that adapts to users' ever-changing environments and needs.

One key aspect to address is the different levels of immersiveness in the mixed reality spectrum. Users may interact with fully immersive environment or minimal augmented reality user interfaces, and the context-aware UI must intelligently adapt to provide optimal interactions in each scenario.

Moreover, the composition and decoration of UI elements play a vital role in enhancing usability. The context-aware UI design should carefully consider the positioning, size, and visual cues of UI elements to enable efficient interactions regardless of the user's surroundings or activities. Whether users are in low-light environments, standing, sitting, or moving through crowded spaces, the UI should dynamically adjust to remain accessible and intuitive.

Considering the varying moods and goals of users further emphasizes the need for a flexible approach. The context-aware dynamic UI should intelligently respond to the specific context and user intentions, providing tailored experiences for work-related tasks or entertainment, ensuring that users have the necessary functionality at their fingertips when they need it.

The other limitation of study is that this study is conducted in 2022, and the devices such as Quest 3 that provides more than 100 FOV and colorful passthrough was not